\documentclass[10pt]{article}
\usepackage{a4wide,graphics,amssymb}

\newcommand{\be}{\begin{equation}}
\newcommand{\ee}{\end{equation}}
\newcommand{\bea}{\begin{eqnarray}}
\newcommand{\eea}{\end{eqnarray}}
\newcommand{\bref}[1]{(\ref{#1})}
\newcommand{\MP}{\ensuremath{M_\mathrm{Pl}}}
\newcommand{\TeV}{\mathrm{TeV}}

\newcommand{\Leff}{\ensuremath{\Lambda_\mathrm{eff}}}
\newcommand{\GB}{\mathcal{L}_\mathrm{GB}}
\newcommand{\alt}{\lesssim}
\newcommand{\agt}{\gtrsim}

\begin{document}

\title{\bf Avoidance of Naked Singularities in Dilatonic Brane World
Scenarios with a Gauss-Bonnet Term}
\author{Pierre Bin\'etruy\thanks{E-mail: 
Pierre.Binetruy@th.u-psud.fr}~$^{1,2}$,
Christos Charmousis\thanks{E-mail: Christos.Charmousis@th.u-psud.fr}~$^1$, \\
Stephen Davis\thanks{E-mail: Stephen.Davis@th.u-psud.fr}~$^1$ and 
Jean-Fran\c{c}ois Dufaux\thanks{E-mail: Dufaux@th.u-psud.fr}~$^1$ \\
\\
${}^1$ \em LPT, Universit\'e de Paris-Sud, B\^at. 210,
 \\ \em 91405 Orsay CEDEX, France \\
${}^2$\em F\'ed\'eration de Recherche APC, Universit\'e Paris 7} 
\date{\today}

\maketitle

\hfill {LPT-ORSAY-02-61, {\tt hep-th/0206089}}

\begin{abstract}
We consider, in 5 dimensions,  the low energy effective action induced
by heterotic string theory including the leading stringy correction
of order $\alpha'$. In the presence of a
single positive tension flat
brane, and an infinite extra dimension, we present a particular class of
solutions with finite 4-dimensional Planck scale and no naked
singularity. A ``self-tuning'' mechanism for relaxing the cosmological
constant on the brane,  without a drastic fine tuning of parameters, is
discussed in this context. Our solutions are distinct from the
standard self-tuning solutions discussed in the context of vanishing quantum
corrections in $\alpha'$, and become singular in this limit.
\end{abstract}

\section{Introduction} 

There has been considerable interest in recent years in investigating
the cosmological aspects of extra dimension (toy) models where matter
and fundamental gauge interactions are localized on a four-dimensional
spacetime surface or 3-brane~\cite{LOSW,LOW,add,BDL}. Interest in such a setup
comes from string theory but it is also inspired by some more
phenomenological motivations. In
particular, it was pointed out that the large hierarchy between the
Standard Model and  the Planck scale  could be redefined for an observer
living on a (negative tension) flat domain wall (or 3-brane), in terms
of the size of the extra dimension~\cite{rs1}. It was also 
realized~\cite{RSA,Rubi,EXO,rs2} that
if a non-compact 5 dimensional spacetime is sufficiently warped, then
four-dimensional gravity is recovered for  observers on a positive
tension flat brane.
 
In both models, a fine
tuned relation between the bulk curvature and the brane tension 
has to be specified in order to switch off the effective cosmological
constant, \Leff, on the brane. Without any specific dynamical
mechanism to justify it, this fine-tuning may be seen as a new version of the 
cosmological constant problem in this context. 

This was the starting point for a series of efforts aimed at resolving
this fine tuning problem by means of a dynamical mechanism, called
``self-tuning''~\cite{ADKS,selftune}. 
A static scalar field, which loosely
models the dilaton and moduli fields of string theory, is added to the
bulk. The extra degree of freedom  is then used to ensure the
existence of a solution of the dynamical equations with a
zero effective brane cosmological constant, whatever the value of the
brane tension. Unfortunately such scalar field solutions generically
suffer from either naked singularities at a finite proper distance from
the brane or badly defined effective
four-dimensional gravity (see for example~\cite{sing}). 
One may screen the naked singularity by including a
second negative tension brane. However this introduces a fine-tuning of
the brane tensions, and it has been  rightfully argued that this is yet again
a  rephrasing, and not a solution, of the cosmological constant
problem~\cite{steph}. 

All of these efforts are of course taking gravity as classical.
However, low energy effective string actions do include the leading order
string quantum gravity corrections~\cite{gross,metsaev}. 
It is widely accepted that the Gauss-Bonnet
term{\footnote{To be precise, string theory, using scattering amplitude
or string worldsheet $\beta$ function calculations, 
predicts only the Riemann squared term, the two remaining terms are put in
by hand in order to ensure the good definition of the resulting
five-dimensional gravity theory, i.e.\ second order derivative equations
of motion~\cite{lovelock} and no gravity ghosts~\cite{zwiebach}.
Concerning the presence of ghosts
however, it was rightly pointed out
by Gross and Sloan~\cite{gross} that they 
appear at high momenta and thus are beyond the
validity of the effective action.}},
\be
\GB=R^2 - 4 R_{ab}R^{ab} + R^{abcd} R_{abcd} \ ,
\label{LGB}
\ee
is the leading order quantum gravity 
correction in heterotic closed string theory at
tree level~\cite{gross,metsaev}. Several
earlier works have been interested in uncovering properties
of 5 dimensional Gauss-Bonnet gravity~\cite{gbgrav} and its consequences in
brane world scenarios~\cite{gbcos,dufaux,greeks,greek2}.

Low and Zee~\cite{lowzee} have thus considered braneworld scenarios with a
5-dimensional bulk action which includes the Gauss-Bonnet term as follows 
$$
S_\mathrm{bulk} = \frac{M_s^3}{2} \int d^5x  \sqrt{-g}  
\left\{ R - \zeta(\nabla \phi)^2+
 \alpha \GB - 2\Lambda e^{\zeta \phi} \right\} \ .
$$
The solution they found did not help in solving the problems of 
the self-tuning scenario. However, as stressed by 
Mavromatos and Rizos~\cite{greeks},
an action motivated by string theory would not
generically be of that form. The reason is simple: in the string frame
the Einstein-Hilbert and Gauss-Bonnet terms have the same dilaton
dependence~\cite{FT} in the effective action. Thus, as can be seen from
the string tree level effective action for the massless boson
sector~\cite{myers},
$$
S^\mathrm{(string)}_\mathrm{bulk} = \frac{M_s^{D-2}}{2}\int d^Dx  
\sqrt{-\bar g}  e^{-2\Phi} \left\{ \bar R +4(\nabla \Phi)^2+
 \lambda_0 \alpha'  \left[ \bar \GB + \cdots \right] 
- \frac{2(D-10)}{3\alpha'} + O(\alpha'^2) \right\} \ ,
$$
where ``$\cdots$'' denotes other four derivative
terms~\cite{gross,metsaev}. The contribution to the cosmological constant
vanishes for a critical ($D=10$) string~\cite{myers}. In this 
case, one has $\lambda_0 = 1/4, 1/8$, and $0$ respectively for the
case of the bosonic, heterotic and type II superstring
theories~\cite{metsaev}. The string coupling is simply
\be
g_s^2 = e^{2 \Phi} \ .
\ee
Applying a conformal transformation 
($\bar  g_{\mu\nu} = e^{\zeta \Phi} g_{\mu\nu}$),we obtain the action in
the Einstein (physical) frame, where the Gauss-Bonnet term 
couples to the dilaton~\cite{gross,metsaev}
\begin{eqnarray}
&& S^\mathrm{(Einstein)}_\mathrm{bulk} = \frac{M_s^{D-2}}{2}
\int d^Dx  \sqrt{-g}  \left\{ R
- \zeta (\nabla \Phi)^2
\right. \nonumber \\ && \hspace{1in} \left. {}
+ \lambda_0 \alpha'  e^{-\zeta\Phi}\left[\GB + \zeta^2 {D-4 \over D-2} (\nabla
\Phi)^4 \right]
- \frac{2(D-10)}{3\alpha'} e^{\zeta\Phi} + O(\alpha'^2)  \right\} \ .
\label{Daction}
\end{eqnarray}
where $\zeta = 4/(D-2)$. As emphasized in \cite{gross,metsaev}, this
action is unique up to field redefinitions which preserve the symmetries
of the theory (i.e.\ general covariance, since we have not considered here
the antisymmetric tensor nor the gauge fields). Choosing the
Gauss-Bonnet combination for the terms quadratic in the curvature requires
that we also consider the term quartic in $\nabla \Phi$, as given
in (\ref{Daction}).

This is the general form of the action that we will consider in this paper. 
We shall show that there exists a
simple solution which, in the presence of a single positive
tension brane, has the following properties: 
\begin{itemize}
\item{It is free of naked singularities, so the curvature tensor is
everywhere regular.} 
\item{The volume element is finite i.e.\ we can define an effective 
4-dimensional Planck mass.}
\item{The solution exists for any value of the brane tension; however it 
does not avoid fine-tuning.}
\item{The solution is singular in the extra
dimension if we take the limit $\alpha' \rightarrow 0$ i.e.\ the solution 
does {\it not}  reduce to any solution of ordinary Einstein
gravity. Previous no-go theorems for
self-tuning~\cite{selftune,Weinberg,in,Cline} may be bypassed here.} 
\item{Higher order string
corrections in $\alpha'$ 
become important as we approach the brane. This one would
expect since Standard Model matter is highly localised there. Quantum
loop corrections on the other hand can be generically small.} 
\end{itemize}

\section{Action with Gauss-Bonnet Terms}

As a five-dimensional toy model, we will consider the effective (low
energy) action given in the Introduction, assuming that it describes a
non-critical string in 5 dimensions~\cite{myers}. In the Einstein frame
the action takes the general form 
\be
S_\mathrm{bulk} = \frac{M_s^3}{2} \int d^4x dz  \sqrt{-g}  
\left\{ R - \zeta (\nabla \phi)^2+  \alpha f(\phi) \left[ \GB + c_2
(\nabla \phi)^4 \right]- 2\Lambda e^{\zeta \phi} \right\} \ ,
\label{action5}
\ee
where $\Lambda$ is the 5-dimensional cosmological constant (in the
string frame). 
The constant $\alpha$ is proportional to $\alpha'\sim M_s^{-2}$ [see
eq.~\bref{Daction}]. In the case of our $D=5$ non-critical string model
we have $\zeta = 4/3$ and $c_2=16/27$. Typically 
\be
f(\phi) = e^{-\zeta \phi} \ .
\ee

We also consider Standard Model matter confined on a positive tension
brane which couples to the bulk dilaton field via the function $\lambda(\phi)$,
\be
S_\mathrm{brane} =  - M_s^3 \int d^4x \sqrt{-h} \, \lambda(\phi) \ ,
\label{action4} 
\ee
with $h_{ab}$ the induced metric on the brane. The brane is taken 
to be static and positioned at $z=0$.
The form of $\lambda$
depends on the type of brane which is considered. For example, in the
string frame, we expect to have 
\be
S^\mathrm{(string)}_\mathrm{brane} =  
-  \int d^4x \sqrt{-\bar h} \, T e^{-\beta \phi} \ ,
\label{action4string} 
\ee
where $T$ is the brane tension (corresponding to the vacuum energy on
the brane) and $\beta = 1$ for a Dirichlet brane and $2$ for a
Neveu-Schwarz brane. The
corresponding action is found in the Einstein frame by performing the 
conformal transformation discussed in the previous section (i.e. $\bar
h_{ab} = e^{\zeta \phi} h_{ab}$). We thus take for the function 
$\lambda(\phi)$
\be
\lambda(\phi) = \frac{T}{M_s^3} e^{\chi \phi} \ .
\label{confcoupl} 
\ee
With our choice $D=5$, the Dirichlet brane corresponds to $\chi=5/3$ and
the Neveu-Schwarz brane to $\chi=2/3$.

Varying the total action $S_\mathrm{bulk} + S_\mathrm{brane}$  
with respect to the 5-dimensional metric, $g_{ab}$, gives the field equation
\bea
&& 
G_{ab} + 2\alpha  \left[ f(\phi) H_{ab}  -2
P_{eabc} \nabla^e \nabla^c f(\phi) \right]
-\zeta \nabla_{\! a} \phi \nabla_{\! b} \phi
+ \frac{\zeta}{2} g_{ab} (\nabla \phi)^2
\nonumber \\ && \hspace{0.5in} {}
+\alpha f(\phi) c_2 
\left[2(\nabla \phi)^2 \nabla_{\! a} \phi
\nabla_{\! b} \phi -\frac{1}{2}g_{ab}(\nabla \phi)^4 \right]
+  g_{ab} \Lambda e^{\zeta \phi} = - \lambda(\phi) \,
\frac{\sqrt{-h}}{\sqrt{-g}} \delta(z) h_{ab} \ ,
\eea
where 
\be 
P_{c ab e} = R_{c ab e} + R_{ab} g_{ec} + R_{ec} g_{ab}
- R_{b c} g_{ae} - R_{ae} g_{bc}  + 
\frac{1}{2} R(g_{ae}g_{bc} - g_{ab}g_{ce})
\ee
is the divergence free part of the Riemann tensor, and 
\be
H_{ab} = \left( R R_{ab} - 2 R_{ac}R^c{}_b
-2 R^{c d}  R_{acbd} + R_a{}^{c d e} R_{b c d e}\right)
-\frac{1}{4} g_{ab} \GB
\ee
is the Lovelock tensor~\cite{lovelock}.

Varying with respect to $\phi$ gives
\bea
&&
2 \zeta \nabla^2 \phi + \alpha f'(\phi) [ \GB -3 c_2 (\nabla \phi)^4]
- 4\alpha c_2 f(\phi) \left[(\nabla \phi)^2 \nabla^2 \phi 
+ 2 \nabla^a\phi \nabla^c\phi \nabla_{\! a}\nabla_{\! c}\phi \right] 
- 2\zeta \Lambda e^{\zeta \phi} 
\nonumber \\ && \hspace{1.5in} {}
= 2\frac{d \lambda}{d \phi} \, \frac{\sqrt{-h}}{\sqrt{-g}} \delta(z) \ .
\eea

We are interested in Poincar\'e invariant solutions for the 4
dimensional spacetime, 
so in 5 dimensions we consider the general conformally flat spacetime, 
\be
ds^2 = e^{2A(z)} \eta_{\mu\nu} dx^\mu dx^\nu + dz^2
\label{ansatz1} 
\ee
and take $\phi$ to be a function of $z$ only.
The field equations now reduce to
\bea
&&
6A'' + 12A'^2+ \zeta \phi'^2 + 2\Lambda e^{\zeta \phi} 
+ 2 \lambda(\phi) \delta(z)
\nonumber \\ && \hspace{0.2in} {}
- \alpha e^{-\zeta \phi}\left[c_2 \phi'^4 +
24 A'(A''A' +A'^3 -2 \zeta A''\phi' -\zeta\phi''A' -3\zeta A'^2\phi' +
\zeta^2A'\phi'^2)\right] = 0
\label{eq1}
\eea
\be
12A'^2- \zeta \phi'^2 +2 \Lambda  e^{\zeta \phi} 
+ \alpha e^{-\zeta \phi}  \left[3c_2 \phi'^4 
-24A'^3(A'-4\zeta\phi')\right]  = 0
\label{eq5}
\ee
\bea
&& 
2\zeta\phi'' + 8\zeta A'\phi' - 2\Lambda \zeta e^{\zeta \phi} 
- 2{d\lambda(\phi)\over d\phi} \delta(z)
\nonumber \\ && \hspace{0.3in} {}
- \alpha e^{-\zeta \phi} \left[
 c_2\phi'^2(12\phi''+16A'\phi'-3\zeta \phi'^2)  
+24 \zeta A'^2(4A''+5A'^2)\right]  = 0 \ ,
\label{eqp}
\eea
where the prime denotes differentiation with
respect to $z$. Due to the generalized Bianchi identities, only two of these
equations are independent. 

By taking an appropriate combination of
eqs.~(\ref{eq1}--\ref{eqp}) we obtain in the bulk, after integration, 
the first order expression
\be
e^{4A}\left( 2\zeta \phi' + 3\zeta A' - 4\alpha e^{-\zeta \phi}
[9\zeta A'^3 - 3\zeta^2 A'^2 \phi' + c_2 \phi'^3]\right) = \mathcal{C} \ .
\label{ic}
\ee
Together with \bref{eq5}, this allows the general solution of the field
equations to be found with relative ease, considering the complicated
action.

\section{Bulk Solutions}

For the sake of clarity we review the self-tuning
solutions for $\alpha=0$. Here we shall take $\zeta=4/3$ and $\Lambda=0$,
as this suffices to illustrate our point. For a full detailed
discussion see ref.~\cite{selftune}. Assuming a $Z_2$-symmetric bulk, the
general solution in this case is,
\be
\label{st}
A(z) = A_0 + \frac{1}{4} \ln\left(1+ \frac{|z|}{z_*} \right) \ , \ \ 
\phi(z) = \phi_0 -  \frac{3}{4}\ln\left( 1+ \frac{|z|}{z_*} \right) \ .
\ee
The effective four-dimensional Planck mass is
\be
\MP^2 =  M_s^3 \int_0^{z_c} dz \, e^{2A} \ ,
\label{MP0}
\ee
where $z_c$ is the maximum value of $z$, so $z_c = \infty$ if $z_* > 0$,
and $z_c=|z_*|$ if $z_* <0$. For $z_*>0$ it is obvious that \MP\ is never
finite. Alternatively if we choose $z_*<0$, \MP\ is finite, but the
curvature diverges as $z \to z_*$. In a nutshell we have either a
singular solution with localised gravity or a regular solution with
non-localised gravity. The reader will note that the constant appearing
in (\ref{ic}) is {\it not} zero. 

We will now investigate the case of interest, $\alpha > 0$ (as dictated
by string theory). 
Inspection of the field equations (\ref{eq1}--\ref{eqp}), 
shows that a similar anzatz to \bref{st},
\be
A(z) = A_0 + x \ln\left(1+ \frac{|z|}{z_*} \right) \ , \ \ 
\phi(z) = \phi_0 -  \frac{2}{\zeta}\ln\left(1+\frac{|z|}{z_*} \right) \ ,
\label{ansatz2}
\ee
reduces them to algebraic equations~\cite{greeks}. There are two fundamental
differences however. First, unlike \bref{st}, 
this is not the general solution to (\ref{eq1}--\ref{eqp}). Secondly 
\bref{ic} shows that in this case $\mathcal{C}=0$. Therefore the presence of 
$\alpha$ corrections has actually opened up new branches 
of solutions, of a form similar to  (\ref{st}) but, as we
will see, with
different properties. Actually there is another class of solutions 
of the form \bref{ansatz2}. Taking $\mathcal{C}\neq 0$, we are led to $x=1/4$.
There are also non-logarithmic solutions to the field equations.
This class of solutions gives a rather complicated
functional form for $A$ and $\phi$ and will be discussed in detail
in future work. We set the constant $A_0$ to zero, as it simply
corresponds to a rescaling of coordinates.

Taking the constant in \bref{ic} to be zero and using \bref{eq5}, we find $x$
as a function of $\Lambda$. It is given implicitly by the equation
\bea
&&
\Lambda =-(-45\zeta^5x^5+36\zeta^5x^4-12\zeta^4x^4-78\zeta^4x^3+12
\zeta^4x^2+48\zeta c_2 x^2-18c_2 \zeta x +8 c_2) 
\nonumber \\ && \hspace{1in} {}\times
\frac{\zeta^2(4-3\zeta x)}{4\alpha (8 c_2 - 9 \zeta^4x^3 - 6\zeta^4x^2)^2} \ .
\label{lameq}
\eea
We also obtain $z_*$ in terms of $x$ and $\phi_0$ 
\be
\frac{4\alpha}{z_*^2} =  e^{\zeta \phi_0} \frac{(4-3\zeta x)\zeta^3}{8 c_2 
- 9 \zeta^4 x^3- 6\zeta^4 x^2}  \ .
\label{p0eq}
\ee
We note here that $z_*$ is  not an integration constant, contrary to the 
case of the self-tuning solution with $\alpha=0$. On the other hand, 
$\phi_0$ is not determined by the bulk solution and is an integration 
constant. Since $\phi_0$ must be real, not all values of $x$ are permitted.

By integrating over the bulk coordinate an effective four-dimensional
action can be obtained for an observer on the brane. The effective
Planck mass with $O(\alpha)$ corrections is then obtained~\cite{greek2}
\be
\MP^2 = M_s^3 \int_0^{z_c} dz e^{2 A(z)} \left(1 + 4\alpha
e^{-\zeta\phi(z)}(3A'^2(z) +2 A''(z)]\right) \propto 
\left[\frac{z_*}{2x+1} \left(1+\frac{z}{z_*}\right)^{2 x+1}\right]_0^{z_c} \ .
\label{MP1}
\ee 
The integral is qualitatively similar to the $\alpha\neq 0$ case
\bref{MP0}. If $x \geq -1/2$, then finite \MP\ requires $z_* <0$ 
(this case was very recently discussed in~\cite{greek2}) and compact
proper distance, while a
naked singularity-free spacetime requires $z_* > 0$. This is similar to
the original $\alpha=0$ self-tuning model. However the important
difference is, that if we can find
solutions with $x <-1/2$, we will have a non-compact
spacetime with no singularities and \MP\ finite, for $z_* > 0$. In this case,
\be
\MP^2 =  M_s^3 \frac{4 z_*}{|2x+1|} 
\, \frac{4 c_2 -9 x^3 \zeta^4+6\zeta^3 x^2-4\zeta^3 x}
{8c_2 - 9 \zeta^4 x^3- 6\zeta^4x^2} \ ,
\label{MP}
\ee 
which determines the Planck mass in terms of the constant $z_*$. The
Ricci scalar, and other curvature invariants are finite for all $z$,
so this range of parameters will give singularity free solutions.  
The only other
range of parameters ($z_* <0$, $x <-1/2$) has infinite \MP\ and singularities.

The appearance of a new  singularity-free solution, once the
Gauss-Bonnet term coupled to a scalar  field is included, may be compared with 
a similar  observation made in a
work by Antoniadis, Rizos and Tamvakis~\cite{ART} in the context of
timelike singularity (in 4 dimensions).

Henceforth we will concentrate on the case  $x<-1/2$ and
$z_*>0$. Note that the string coupling $g_s = e^\phi$
decreases away from the brane. Thus keeping $e^{\phi_0}$ small will 
guarantee validity of perturbative string loop corrections. In
that case any solution will be stable to higher order quantum loop corrections.

\begin{figure}
\begin{center}
\includegraphics{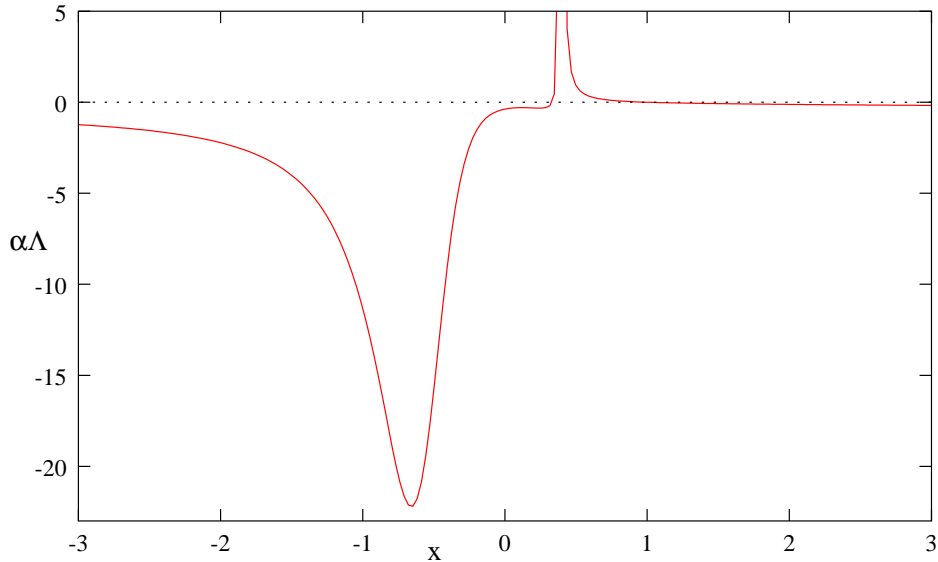}
\end{center}
\caption{Variation of $\alpha \Lambda$ with respect to $x$, for $\zeta = 4/3$
and $c_2=16/27$.} 
\label{fig1}
\end{figure}

As a definite example, we will now take $\zeta=4/3$ and $c_2=16/27$. 
Equation~\bref{p0eq} has real solutions when $x>1$ or $x \alt 0.4$. Note
that if we had neglected the fourth-derivative dilaton term in the
action~\bref{action5} the acceptable range of $x$ would be reduced.  The
number of solutions to \bref{lameq} varies with $\Lambda$, as can be
seen from figure~\ref{fig1}. For $\alpha \Lambda \alt -22.2$,  there are
no solutions at all. For $-22.2 \alt \alpha \Lambda < -63/4$, there are two
solutions with $x<-1/2$. For $-63/4 < \alpha \Lambda <-5/12$ there is
one solution with $x<-1/2$ and one with $x>-1/2$. For larger values of
$\alpha \Lambda$ there are between one and four solutions, all with 
$x >-1/2$. Hence the only acceptable solutions must have a negative bulk
cosmological constant.

If we take the limit of $\alpha \to 0$, with $z_*$ and $\phi_0$
remaining finite, the $x < -1/2$ branch
of solutions to \bref{lameq} will disappear. We will be left with only
$x>1$ solutions, which have singularities or infinite \MP.

Alternatively we can rescale the dimensionful integration constant
$z_*$, defining $\bar{z}_*=z_*/ \sqrt{\alpha}$. Then going back to 
(\ref{ansatz2}) we see that the $\alpha=0$ limit is badly
behaved (Direct integration of the field equations gives
an imaginary string coupling). This is not surprising since as 
eq.~\bref{ic} already
tells us, $\alpha=0$ and $\mathcal{C}=0$ are incompatible.

\section{``Self-Tuning'' Mechanism on the Brane}

We now move on to the junction conditions which will relate our bulk
parameters to the physical brane tension.
We assume $Z_2$ reflection symmetry about $z=0${\footnote{The
breaking of $Z_2$ symmetry could result in additional stability
concerns since a scalar degree of freedom, much like the radion for 
a two brane configuration, would appear corresponding to
brane fluctuations.}}. 
The junction conditions at $z=0$ obtained from \bref{eq1} and \bref{eqp}
are~\cite{greek2}
\bea
\left[3A' + 4 \alpha e^{-\zeta\phi}
 (3\zeta\phi'A'^2 - A'^3)\right]_-^+ = - \lambda \ , 
 \\ \left[\zeta \phi' - 2 \alpha e^{-\zeta\phi} (c_2\phi'^3
+ 8\zeta A'^3 )\right]_-^+ = \frac{d\lambda}{d\phi}  \ .
\eea 
Substituting in our solution we find that $\chi =\zeta/2$. Thus, for 
$\zeta=4/3$, $\chi=2/3$ which is is the
value we would expect from a Neveu-Schwarz brane. This is thus
absolutely consistent with our original choice of a heterotic string
model.

We find that the brane tension $T$ satisfies
\be
T = M_s^3 \  e^{-\zeta\phi_0/2}
\frac{(-8x) 
(6c_2 - 6\zeta^4x^3 - \zeta^3 x^2 - 6\zeta^3 x)}
{z_*(8 c_2 - 9\zeta^4x^3 -6\zeta^4 x^2)} \ .
\label{Teq}
\ee
However, as noticed earlier, $z_*$ is not an integration constant\footnote{This
should be contrasted with the self-tuning solution with $\alpha = 0$, where
$\chi = 1$ and
$$T =  M_s^3 \  e^{-\phi_0}{3 \over 2 (-z_*)}$$
and both $z_*$ and $\phi_0$ are integration constants.}
it is determined in terms of the parameters of the theory through \bref{p0eq}.
Thus we have
\be
T = \frac{M_s^3}{\sqrt{\alpha}} \  (-4x) \
\frac{ (6c_2 - 6\zeta^4x^3 - \zeta^3 x^2 - 6\zeta^3 x)(4-3\zeta x)^{1/2}
\zeta^{3/2}}
{(8 c_2 - 9\zeta^4x^3 -6\zeta^4 x^2)^{3/2}} \ ,
\label{FT}
\ee
which amounts to a fine-tuned relation among the parameters of the theory 
since $x$ is related to $\alpha \Lambda$ through \bref{lameq}.

We will now take again  $\zeta=4/3$ and $c_2=16/27$ as a specific
example. 
If $z_*>0$ and $x<0$ the
tension is always positive, so our model uses a physical brane. When
$z_*<0$ it has negative tension unless $x>0$~\cite{greek2}. We note
that, on the contrary, in the standard self-tuning solution, we have
an opposite sign for $T$ and $z_*$ (see previous footnote).

Using equations~\bref{p0eq},\bref{MP} and \bref{FT} we can obtain the
string coupling and $M_s$ in terms of \MP\ and $T$. If $(-x) \agt 1$,
which is the case when $\alpha\Lambda$ is of order $-1$, we obtain the
order of magnitude estimates
\be
e^{\phi_0} \sim \frac{M_s^3}{\MP^3} \ ,
\label{gmag}
\ee
\be
T \sim M_s^4 \ .
\ee
If we take $T \sim (1 \TeV)^4 \sim 10^{-60} M_{Pl}^4$, 
we obtain $e^{\phi_0} \sim 10^{-45}$ which is obviously a ridiculously small 
value for the string coupling: we recover the typical  fine-tuning associated 
with the cosmological constant. Of course, in the context of the weakly
heterotic model, we expect $M_s$ to be much closer to the Planck scale.
On the other hand, we see from eq.~\bref{MP} that if $x$ is near to $-1/2$,
we can obtain $e^{\phi_0}\sim 1$ when $M_s \ll \MP$. However this
requires severe fine-tuning of $x$.

\section{Conclusions and outlook}

In this paper we have shown the existence of a ``quantum corrected''
solution to the system with a bulk dilaton conformally coupled to 
matter on the brane, much as in the self-tuning set up. Unlike previous
work 
on the subject our solution is free of naked singularities. It is
expected to localize 4-dimensional gravity on the brane since the
corresponding Planck scale is computed to be finite. The absence of a
naked singularity allows us to study the issue of fine-tuning without
ambiguity. We find that  a fine tuning is
necessary between the brane tension and the bulk vacuum energy (through
our equations \bref{FT} and \bref{lameq}). 

It must be said however that the higher order corrections that we
include do not resolve  the singularity problems of the
``self-tuning'' solutions found in the $O(\alpha'^0)$ theory. Instead
the $O(\alpha')$ terms produce a completely new branch of well behaved
solutions.  Thus our approach differs from other approaches on the
subject in that the solution no longer exists  if the ``quantum
correction'' is set to zero  (see also~\cite{greeks, greek2}). This is a
genuine 5 dimensional $O(\alpha')$ classical solution. 

In the context of string theory and since
we are relying on an effective action approach, the solution is well
defined  if higher order
quantum gravity corrections in $\alpha'^2$~\cite{gross}  or quantum loop
corrections in $g_s=e^{\phi}$ are not dominant. As we showed the string
coupling is generically small in our model becoming of order 1 when
$M_s\sim M_{pl}$. Furthermore it is easy to see that
quantum gravity corrections in powers of $|\alpha' R|$ 
become important as we get closer to the
brane. This is not surprising, since according to Mach's principle,
spacetime is strongly curved where matter is localised. Since we are
crudely modeling our universe by a Dirac distribution of tension $T$ 
our approximation
will inevitably fail close to the brane. At this scale the fine
structure of Standard Model physics becomes important. 
On the geometry or gravity side this is
consistent with the fact that any higher order quantum gravity
corrections in $\alpha'^2$ would destroy the nice properties of
Gauss-Bonnet gravity (see for example~\cite{dufaux} and references
within). Indeed, field equations would depend on higher derivatives in
the metric, and distributional boundary conditions would be ill-defined!
One would inevitably have to consider our brane as having some
non-trivial $\alpha$-dependent thickness. 

Any toy model, such as the one described here, suffers from its own
limitations as an effective theory. For the ansatz we consider, one can
argue that higher order curvature corrections in
$\alpha'^2$~\cite{gross}  will increase the order of the algebraic
equations for $x$. This could give rise to yet more solutions, although
it is also possible that our solutions may disappear. A full solution to
the cosmological constant problem has to rely on a quantum theory of
gravity.

What we feel is important is the realisation that quantum  corrections
to ordinary Einstein gravity can be crucial in higher dimensional
models. Even if one considers them at first order, as we do here, they
are compatible with the principles of relativity theory for $D>4$. Indeed
the
Gauss-Bonnet action has to be included in the Einstein-Hilbert plus
cosmological constant action in order for the gravity theory to be
unique~\cite{lovelock} in 5 dimensions. On the other hand these terms
provide a window to the leading $\alpha'$ gravity string
corrections. Furthermore if there is to be a low string scale (of order
1 TeV) then these $\alpha'$ corrections become more and more
important. 

An interesting direction of study is whether solutions of the type
discussed in this paper exist if we allow a small cosmological constant
on the brane, i.e. if the brane is $dS_4$ or $AdS_4$.
Furthermore, the issue of stability is very important here. 
It is essential that the solution we propose is classicaly stable under
small perturbations.
Such a solution has to be a stable attractor 
for other solutions of the system if we really are to talk of a
higher dimensional resolution of the cosmological constant
problem. Furthermore
localisation of 4-dimensional gravity is required  in order to view this
model as a new approach to 
a string theory realisation of the Randall-Sundrum model~\cite{rs2}.

\section{Acknowledgements}

It is a great pleasure to thank Emilian Dudas, Christophe Grojean, 
Jihad Mourad, and Dani Steer 
for numerous discussions on the subject. We are indebted to 
Jihad Mourad for useful remarks.   

\def\prd{Phys.\ Rev.\ D }
\def\prl{Phys.\ Rev.\ Lett.\ }
\def\npb{Nucl.\ Phys.\ {\bf B}}
\def\pla{Phys.\ Lett.\ {\bf A}}
\def\plb{Phys.\ Lett.\ {\bf B}}
\def\cqg{Class.\ Quant.\ Grav.\ }
\def\jmp{J.\ Math.\ Phys.\ }
\def\rmp{Rev.\ Mod.\ Phys.\ }

\end{document}